\documentclass[aps,prapplied,twocolumn,showpacs,superscriptaddress,groupedaddress]{revtex4-1}
\usepackage{graphicx}  
\usepackage{dcolumn}   
\usepackage{bm}        
\usepackage{amssymb}   
\usepackage{subfigure}

\usepackage[utf8]{inputenc}

\begin{document}

\title{Tailored design of mode-locking dynamics for low-noise frequency comb generation}
\author{Çağrı~Şenel} \affiliation{TÜBİTAK National Metrology Institute (UME), 41470, Kocaeli, Turkey}\affiliation{Department of Physics, Boğaziçi University, 34342, İstanbul, Turkey}
\author{Ramiz~Hamid} \affiliation{TÜBİTAK National Metrology Institute (UME), 41470, Kocaeli, Turkey}
\author{Cihangir~Erdoğan} \affiliation{TÜBİTAK National Metrology Institute (UME), 41470, Kocaeli, Turkey}
\author{Mehmet~Çelik} \affiliation{TÜBİTAK National Metrology Institute (UME), 41470, Kocaeli, Turkey}
\author{Fatih~Ömer~Ilday} \affiliation{Department of Physics, Bilkent University, 06800, Ankara, Turkey}\affiliation{Department of Electrical and Electronics Engineering, Bilkent University, 06800, Ankara, Turkey}\affiliation{UNAM-National Nanotechnology Research Center and Institute of Materials Science and Nanotechnology, Bilkent University, 06800, Ankara, Turkey}
%
%
\vskip 0.25cm
 
\date{\today}

\begin{abstract}
We report on a mode-locked laser design using Yb-doped fiber lasers for low-noise frequency-comb generation. The frequency comb covers the spectral range from $700$ to $1400$ nm. Although this range is more practical for many measurements than that produced by the more commonly used Er-fiber lasers, it has been addressed in only a handful of reports, mainly due to the difficulty of generating a fully coherent supercontinuum at $1\;\mu$m. We overcome this difficulty by a tailored design of the mode-locking dynamics that succeeds in generating energetic 33 fs-long pulses without even using higher-order dispersion compensation, while ensuring that the laser operates with net zero cavity dispersion for low-noise supercontinuum generation. After locking to a Cs atomic clock, this frequency comb is used for absolute-frequency measurements of a Nd:YAG-I$_2$ laser to verify its accuracy by comparison with results from the International Committee for Weights and Measures. After this verification, it is further used to measure the absolute frequency of a 543-nm two-mode stabilized He-Ne laser, which is routinely used for length measurements in our institute, thus verifying its practical utility in metrology applications. The entire setup is built with readily available components for easy duplication by other researchers. 
\end{abstract}

\maketitle

\section{Introduction}

The uses of femtosecond combs \cite{evolving_comb} have diversified from optical clocks \cite{clock_Sr1,clock_Sr2}, optical-frequency metrology, and laser spectroscopy to dimensional metrology \cite{comb_gauge}, astronomical-spectrograph calibration \cite{comb_astro}, time and frequency transfer \cite{comb_time_transfer}, long-range absolute-distance measurements \cite{comb_distance}, surface-profile metrology \cite{comb_surface}, low-noise rf-signal generation \cite{comb_microwave}, and even testing for the variations in physical constants \cite{comb_fine}. To harness fully the benefits of an optical frequency comb, the so-called carrier-envelope offset frequency has to be stabilized, which is most commonly achieved by octave-spanning coherent supercontinuum generation. Octave-spanning coherent supercontinuum generation with photonic crystal fibers (PCFs) usually requires pumping of a PCF with short pulses ($<$50 fs) \cite{SC_review}. Total dispersion of the laser cavity affects the width of the carrier-envelope offset-frequency beat note, and it gets wider as total cavity dispersion increases. Zero cavity dispersion is desirable for this reason \cite{Impact_of_GVD}. The sensitivity of these complex setups and the need to minimize noise coupling to various nonlinear processes for stabilization of a frequency comb complicate long-term operation and limit the measurement times for absolute-frequency measurements. These practical limitations were effectively addressed by the development of combs based on Er-fiber lasers operating at a central wavelength around 1.55 $\mu$m \cite{Er_comb}. 
However, the operating wavelength is not ideal for many applications. First, the optical spectrum generated from these systems typically cover the $1 - 2$ $\mu$m wavelength range, whereas most optical-frequency references are in either the visible range or the near-IR range, requiring an additional frequency-conversion process to link the comb to. In contrast, Yb-fiber lasers surpass Er-fiber lasers in almost every technical metric, such as power, power efficiency, pulse energy, and minimum-achievable pulse duration. Despite these advantages, few frequency combs based on Yb-fiber lasers have been demonstrated \cite{10_W,80_W,375_MHz, offset_free,750_MHz,Li:17,Cingoz:11,Wang:14,PhysRevA.84.011806}, in contrast to hundreds of reports using Er-fiber lasers. Even though excellent results were demonstrated in Ref. \cite{10_W}, their setup relied on specialty components, a specialty PCF, precise dispersion control of even third-order dispersion (TOD), and an overall highly complex architecture. It is probably not surprising that this setup has not been duplicated by any other group. Follow-up studies on Yb-fiber combs have, in contrast, achieved more modest performance, including pulse durations exceeding 100 fs. Furthermore, none of these \cite{10_W,80_W,375_MHz, offset_free,750_MHz,Li:17,Cingoz:11,Wang:14,PhysRevA.84.011806} were used to make absolute frequency measurements or were used in an application after being linked to an absolute-frequency reference.     

Here, we present a specially tailored design of mode-locking dynamics of a Yb-doped fiber laser to meet the conditions for fully coherent and efficient supercontinuum generation. There is a well known trade-off between gain narrowing that accompanies strong amplification and the desire to generate as short pulses as possible. Nevertheless, we aim for the generation of as short pulses as possible (33 fs was achieved), while ensuring the pulse energy was high (4 nJ was achieved). Simultaneously, the cavity has to have zero total dispersion for low-noise supercontinuum generation, which places an additional constraint. In addition to meeting these constraints related to laser physics, having observed that the overwhelming majority of fiber lasers reported in the literature are never duplicated, we meet a further, self-imposed constraint of using only nonspecialty, commonly available components for easy duplication by our colleagues at our institute as well as other researchers elsewhere. Finally, we confirm the accuracy of a Yb-fiber comb through an absolute frequency measurement.

\section{Mode-Locked Oscillator}

\begin{figure}
	\includegraphics[width=0.5\textwidth]{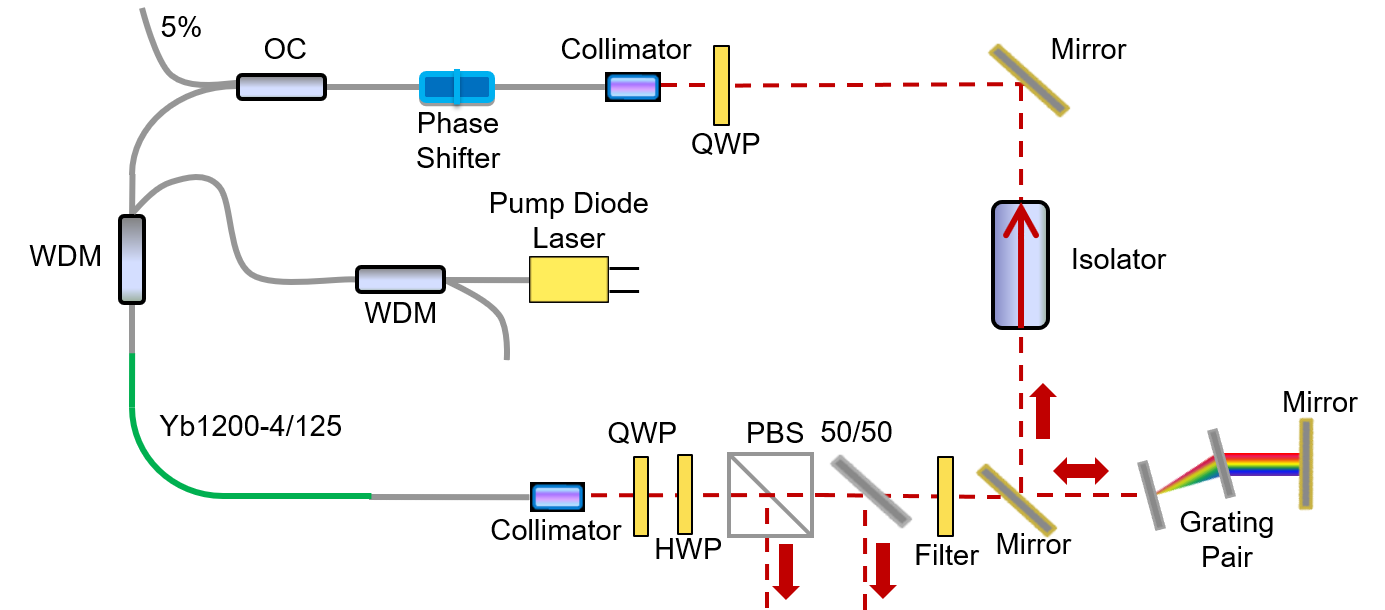}
	\includegraphics[width=0.5\textwidth]{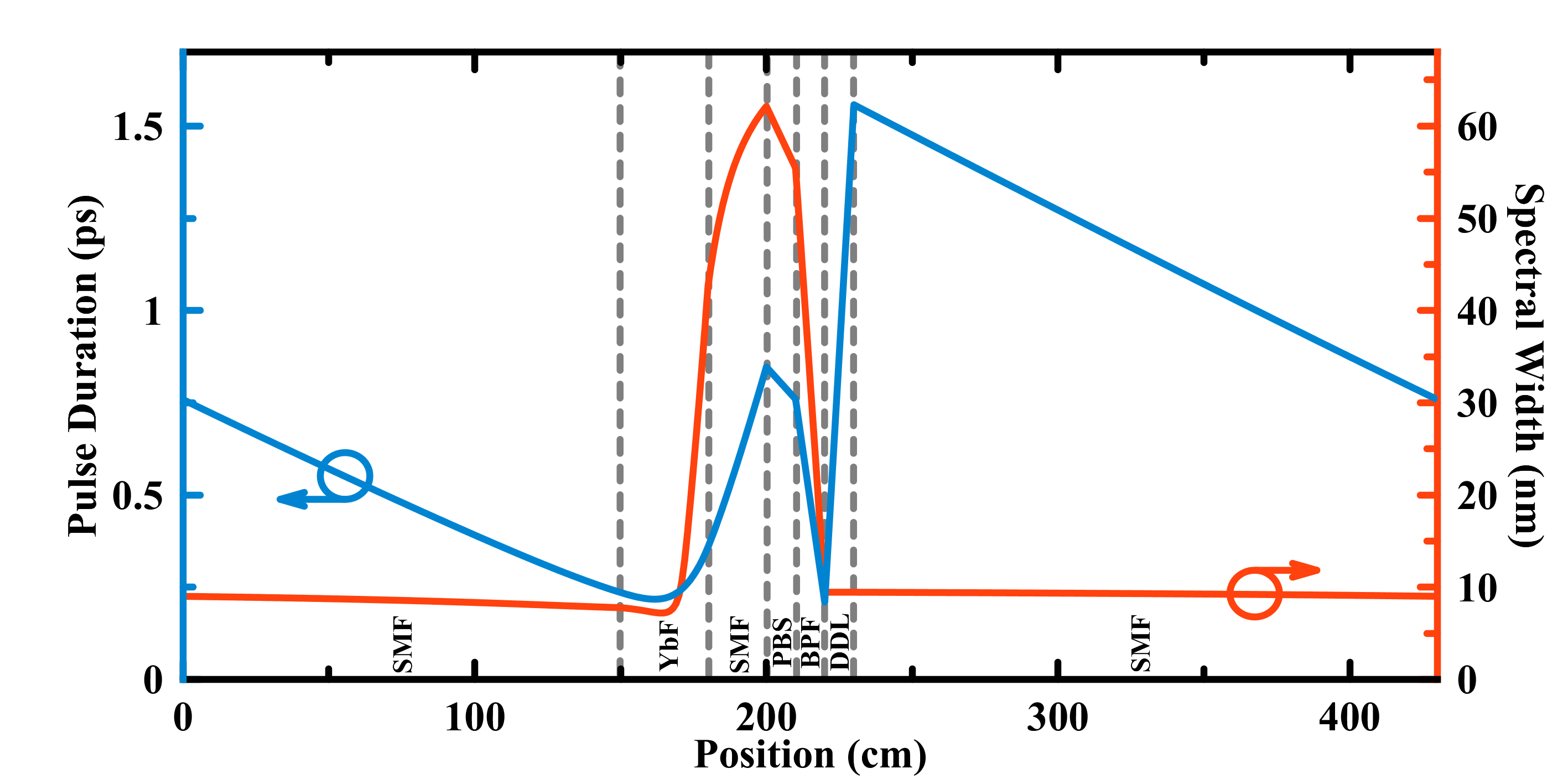}
	\caption{\label{fig:laser} Upper panel: The mode-locked laser oscillator specially designed for fully coherent supercontinuum generation. Lower panel: Simulation results showing the evolution of the pulse inside the laser cavity. BPF, bandpass filter; DDL, diffraction-grating-based delay line; HWP, half-wave plate; OC, output coupler; PBS, polarizing beam splitter; QWP, quarter-wave plate; SMF, single-mode fiber; WDM, wavelength divider multiplexer; YbF, Yb-doped gain fiber.}
\end{figure}

We design the Yb-fiber oscillator through an algorithmic method that makes it possible to simultaneously meet the multiple constraints described above. This method uses an analogy between a mode-locked oscillator and a thermodynamic engine to open up the use of the tools and techniques of nonlinear dynamical systems and nonequilibrium thermodynamics to meet the conditions of the design. We aim to stay below 50 fs and to keep the total dispersion of the cavity at zero, both required to generate a fully coherent supercontinuum \cite{SC_review}. Furthermore, given the typical mode-field diameter of about $4$ $\mu m$ of commercially available PCFs with zero dispersion at $1$ $\mu m$, the pulse energy needs to be at least several nanojoules.  

We employ similariton formation to generate a broad spectrum within a short propagation distance under normal dispersion to minimize partial loss of coherence (broadening of the comb lines) \cite{similariton_Ilday,SC_review}. Then a portion of the pulse is output coupled. To return to initial pulse width and chirp, first a filter is used to reduce the bandwidth, and then anomalous dispersion is applied such that the pulse has negative chirp. Finally, spectral compression is used to return to the initial state. The amount of anomalous dispersion is chosen to set the total dispersion to zero, to which the spectral compression dynamics are insensitive. By adjustment of the filter bandwidth, the degree of spectral compression and its final state is finely controlled.

This conceptual design, which required no tools other pen and paper, is verified and refined with a numerical model of the laser. Simulations also allow more precise values for the various optical parameters to be determined compared with the conceptual design. The fiber section of the laser cavity consists of 350 cm of undoped single-mode fiber (HI1060 in experiments) followed by a segment of 30 cm of Yb-doped gain fiber (Yb1200-4/125 in experiments) and another segment of 20 cm-long HI1060 fiber. The passive fiber is modeled with a nonlinearity coefficient of $5.9\; ({\rm kW}.{\rm m})^{-1}$, group velocity dispersion (GVD) of $24.8\; {\rm fs}^2/{\rm mm}$, and third order dispersion (TOD) of $23.3 \; {\rm fs}^3/{\rm mm}$. The gain fiber has a nonlinearity coefficient of $11.5 \; ({\rm kW}.{\rm m}) ^{-1}$, GVD of $26.2\; {\rm fs}^2/{\rm mm}$, and TOD of $13.4 \; {\rm fs}^3/{\rm mm}$. The effect of the saturable absorption is modeled with a sinusoidal transmission function, which closely approximates nonlinear polarization evolution (NPE) that we use experimentally. The details of the numerical code can be found in Ref. \cite{soliton-similariton}. A 50:50 beam splitter is included, as in experiments, after the polarizers of the NPE. For dispersion control, we use a diffraction-grating based delay line, which is modeled with GVD of $-1510\; {\rm fs}^2/{\rm mm}$, TOD of $2841 \; {\rm fs}^3/{\rm mm}$, and negligible nonlinearity. The bandpass filter is modeled with a Gaussian transmission profile and full width at half maximum (FWHM) of 10 nm. All the optical elements in the simulations correspond to their experimental counterparts and appear in the same order. The experimental setup is depicted in Fig. \ref{fig:laser}. 

\begin{figure}
	\includegraphics[width=0.5\textwidth]{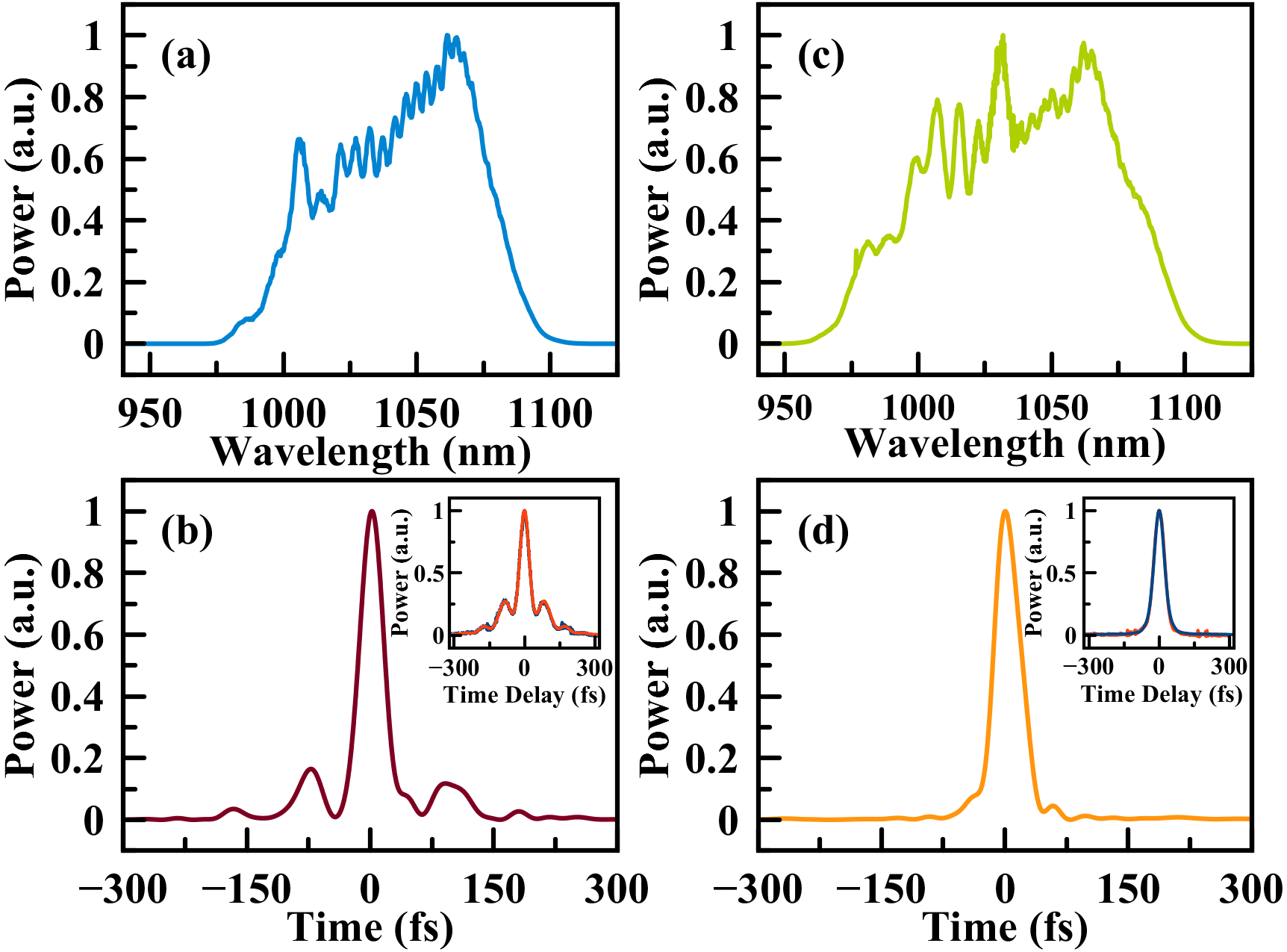}
	\caption{\label{fig:laserampcombined} (a) Measured optical spectrum of the laser output. (b) Inferred pulse shape. The inset shows the measured autocorrelation trace and retrieved autocorrelation trace obtained with the PICASO algorithm. (c) Measured optical spectrum of the amplifier output. (d) Inferred pulse shape for the amplifier output after dechirping. The inset shows the measured autocorrelation trace and retrieved autocorrelation trace obtained with the PICASO algorithm.}
\end{figure}

Simulated evolution of the laser is also shown in Fig. \ref{fig:laser}. The bandpass filter is placed between the polarizing beam splitter (PBS) and diffraction gratings. Pulses enter the fiber segment with large negative chirp and a spectral bandwidth of about 10 nm after the diffraction gratings. Pulses undergo spectral and temporal compression in the fiber until the beginning of the gain fiber. In the gain fiber, they are amplified self-similarly and they continue to evolve in the following segment of single mode fiber. The pulses become slightly shorter and spectrally narrower after the PBS. The filter makes the pulses spectrally narrower and temporally much shorter due to their chirp, restoring the initial state.

We next describe additional details pertaining to the experimental implementation. The gain fiber is pumped by a standard single-mode diode laser (LC962UF76-20R, Oclaro, Inc.) with a wavelength of 976 nm and a maximum output power of 600 mW. A fiber phase shifter (FPS-001, General Photonics, Inc.) is included in the cavity to control the repetition rate. Two quarter-wave plates, one half wave plate and a PBS are placed in the cavity to implement NPE. The PBS is followed by a 50:50 beam splitter for output coupling, and a pair of diffraction gratings of 600 lines/mm (VPH-600-1040-UNP-UHE, Kaiser Optical Systems Inc.) are used in the dispersive delay line. An isolator is also included to ensure unidirectional operation. The bandpass filter is of interference type and centered at 1030 nm (F10-1030.0-4-1.00, CVI Inc.). The laser has a repetition rate of 50 MHz. It is operated at pump powers ranging from 250 mW to 600 mW. The net-cavity dispersion is measured in situ with use of the method proposed by Knox \cite{Knox:92}. To this end, we change the central wavelength of the laser by rotating the filter in the cavity while monitoring the repetition rate of the laser using a frequency counter. We calculate the cavity dispersion by differentiating the pulse repetition rate versus the central frequency of the laser and adjust the grating spacing in the cavity to set the net dispersion to zero. 

\begin{figure}
	\includegraphics[width=0.5\textwidth]{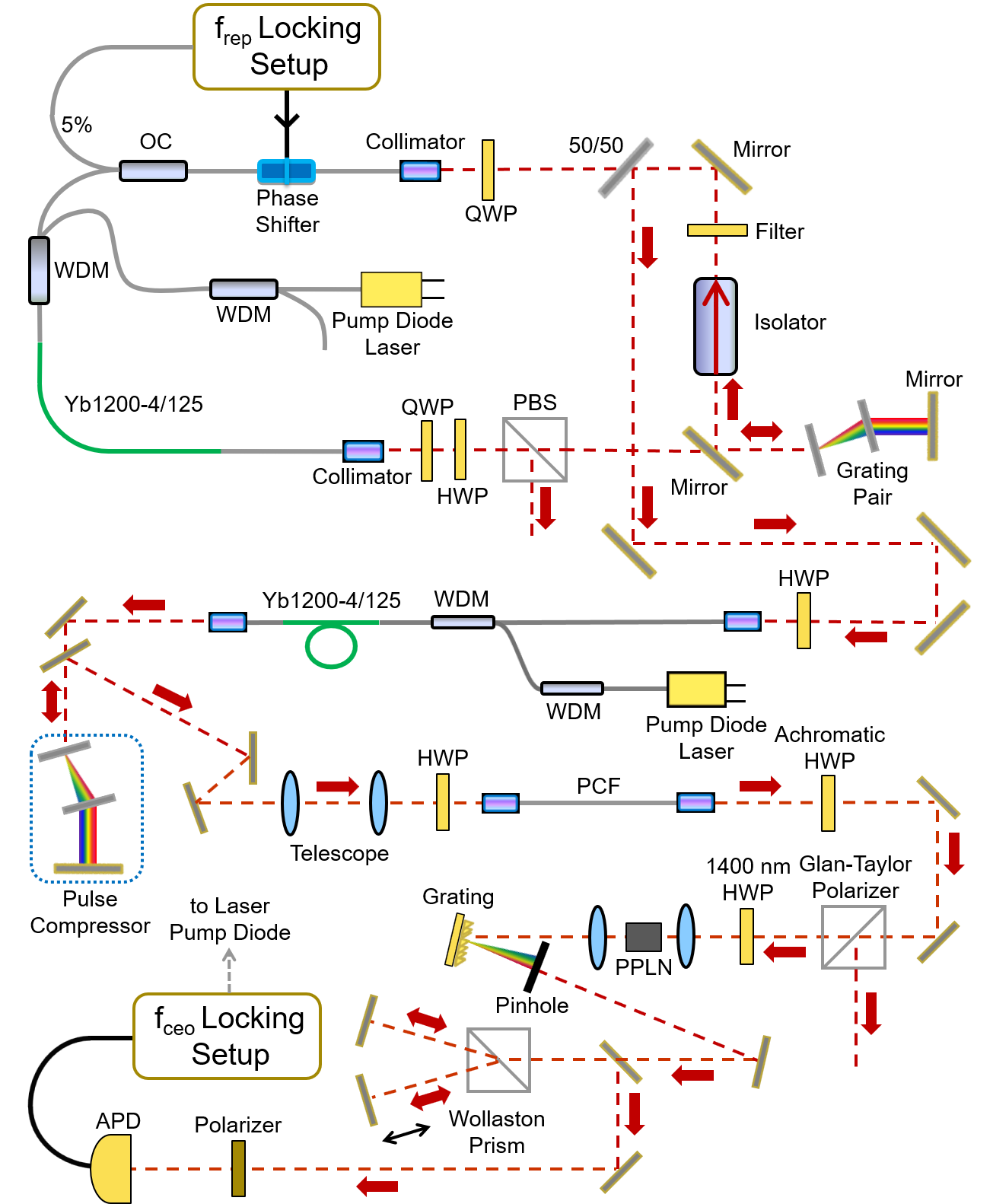}
	\caption{\label{fig:comb} The stabilized frequency comb. APD, avalanche photodetector; HWP, half-wave plate; OC, output coupler; PBS, polarizing beam splitter; PCF, photonic crystal fiber; PPLN, periodically poled lithium niobate; QWP, quarter-wave plate; WDM, wavelength divider multiplexer.}
\end{figure}

It is easy to achieve mode-locking experimentally on adjustment of the waveplates. Intracavity pulse energy before (after) the gain fiber is 60 pJ (4 nJ) for a pump power of 600 mW. The laser's output spectrum is taken from the 50:50 beam splitter. An estimate of the pulse shape is obtained with use of the PICASO algorithm after dechirping of the pulses with diffraction gratings of 600 lines/mm (same model as in cavity). This method of retrieval is not exact and should be interpreted simply as a more refined estimate of the pulse duration than use of only the autocorrelation data. The results are shown in Fig. \ref{fig:laserampcombined}. The FWHM of the pulse is 33 fs. The fluctuations in the output of the laser are characterized by relative-intensity-noise (RIN) and phase-noise measurements. 
RIN measurement method is described in detail in Ref. \cite{RIN}. The integrated RIN from 3 Hz to 250 kHz is measured as 0.017\%.


Phase noise is measured with the $16^{\rm th}$ harmonic of the free-running repetition frequency. The timing jitter calculated with an offset-frequency interval of 1 kHz to 25 MHz is 76 fs. Most of the phase noise stems from the electronic detection and measurement system, and this value should be considered to be an upper bound to the actual phase noise of the laser. Typical timing jitter values measured for near-zero dispersion Yb-doped fiber lasers with optical techniques are less than 1 fs \cite{Yb_noise}.

\begin{figure}
	\includegraphics[width=0.5\textwidth]{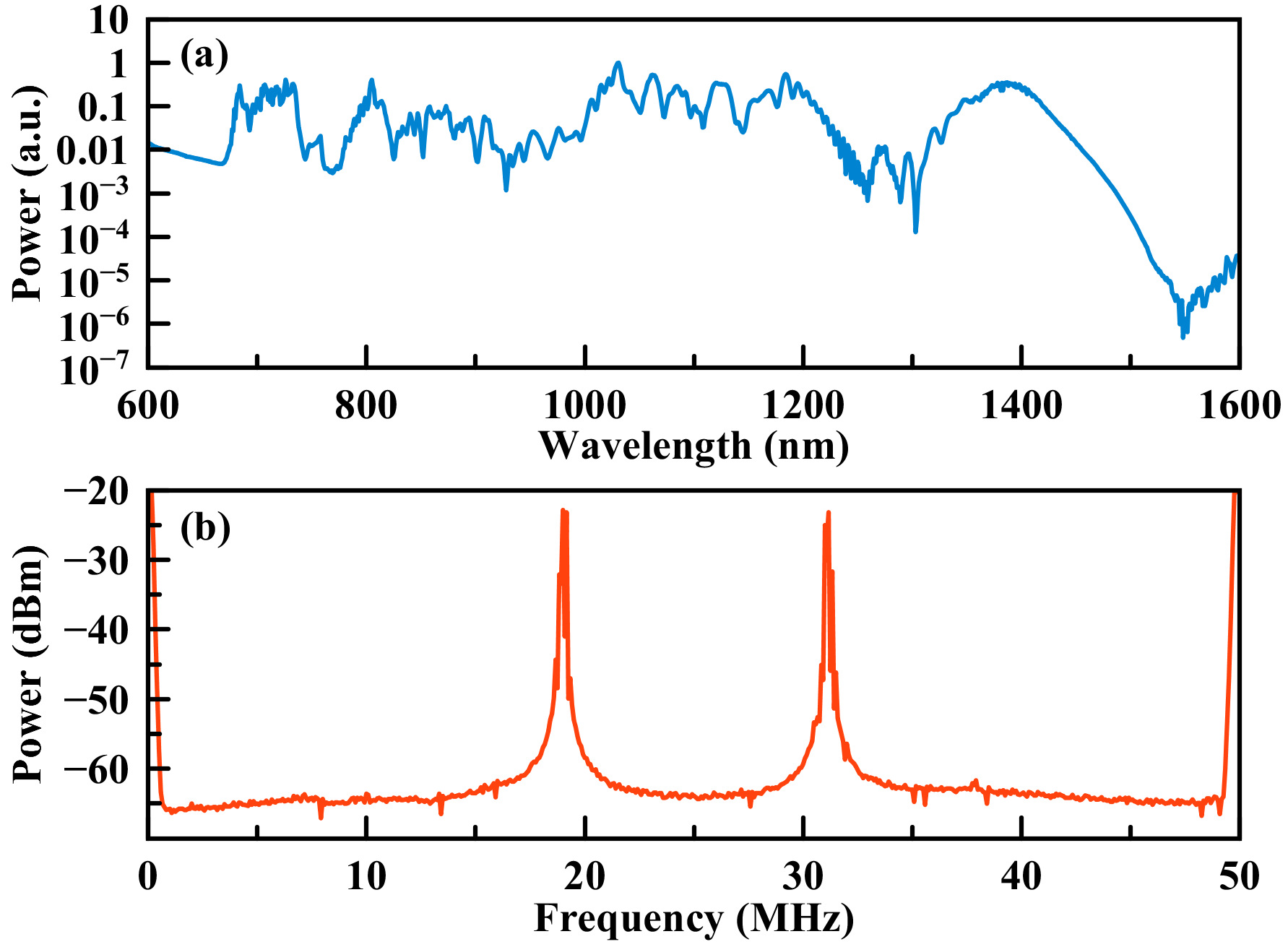}
	\includegraphics[width=0.5\textwidth]{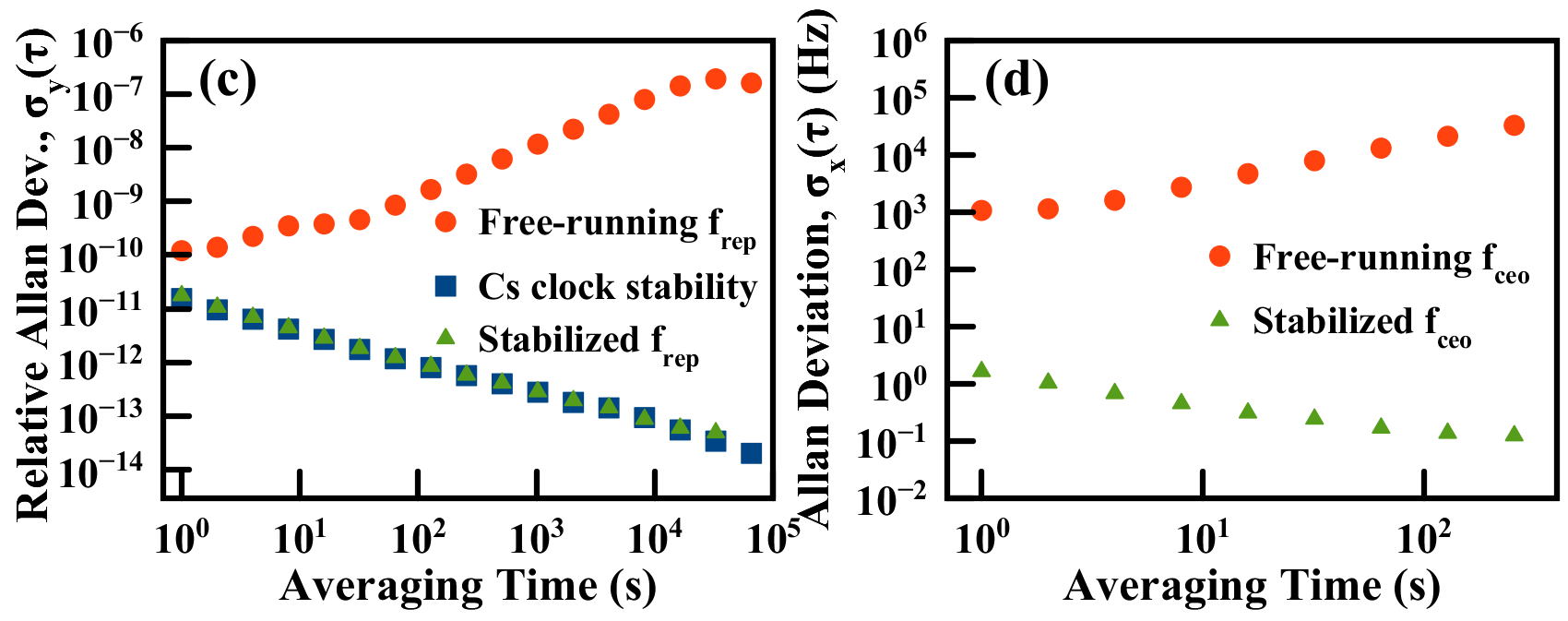}
\caption{\label{fig:SC-fceo} (a) Measured optical spectrum of the generated supercontinuum. (b) Obtained carrier-envelope offset-frequency beat signals. (c) Overlapping Allan deviations for the repetition frequency. (d) Overlapping Allan deviations for the carrier-envelope offset-frequency.}
\end{figure}

\section{Frequency Comb Generation}

After characterization of the laser, we build a stabilized frequency comb. The frequency comb is built in a temperature-stabilized box, which is temperature stabilized by Peltier devices. The box is placed on an optical table without additional vibration isolation. A schematic of the system is shown in Fig. \ref{fig:comb}. Firstly, the filter is placed after the isolator, and the 50:50 beam splitter's location is changed to a point immediately after the filter. A single-stage amplifier, which is identical to the laser's fiber segment, is built and pumped with 700 mW of pump power, implementing a version of in-line fiber amplification demonstrated in Ref. \cite{YDFA}. The beam from the 50:50 beam splitter is coupled into the amplifier, which means that the amplifier and the fiber section of the laser have identical inputs. The output power of the amplifier is 335 mW. The resulting spectrum is shown in Fig. \ref{fig:laserampcombined}. The output of the amplifier is dechirped with a pair of diffraction gratings of 300 lines/mm. The dechirped pulses have FWHM of 35 fs; the estimate for the pulse shape is also shown in Fig. \ref{fig:laserampcombined}.


For supercontinuum generation, 60 mW from the output of the amplifier is coupled into a 30 cm-long piece of PCF (SC-3.7-975, NKT Photonics) with zero-dispersion wavelength of 975 nm to generate a supercontinuum. The supercontinuum spectrum obtained covers the wavelength range from $700$ to $1400$ nm  [Fig. \ref{fig:SC-fceo}(a)]. An $f$-$2f$ interferometer is built in which the second harmonic of 1400 nm is generated with a 1 cm-long periodically poled MgO-doped lithium niobate crystal (HC Photonics) \cite{Wollaston}. Carrier-envelope-offset beat signals are observed by our beating the signals from two outputs of the Wollaston prism on a photodiode [Fig. \ref{fig:SC-fceo}(b)]. The signal-to-noise ratio of the observed beat signal is more than 40 dB. For a resolution bandwidth of 100 kHz, the $-20$-dB width of the free-running beat signal is approximately 400 kHz. The repetition rate and carrier-envelope offset frequency of the laser are locked to the 10-MHz output of a Cs atomic clock. Then their stabilities are characterized by overlapping Allan deviation graphs, as shown in Fig. \ref{fig:SC-fceo}(c) and \ref{fig:SC-fceo}(d). The Allan deviation values for the repetition rate of the free-running laser are $1.2\times 10^{-10}$ for an averaging time of 1 s and $1.6\times 10^{-7}$ for an averaging time of 65536 s. When the repetition rate of the laser is also locked to the Cs atomic clock, its Allan deviation values are $1.8\times 10^{-11}$ for an averaging time of 1 s and $4.9\times 10^{-14}$ for an averaging time of 32768 s.

\begin{figure}
	\includegraphics[width=0.5\textwidth]{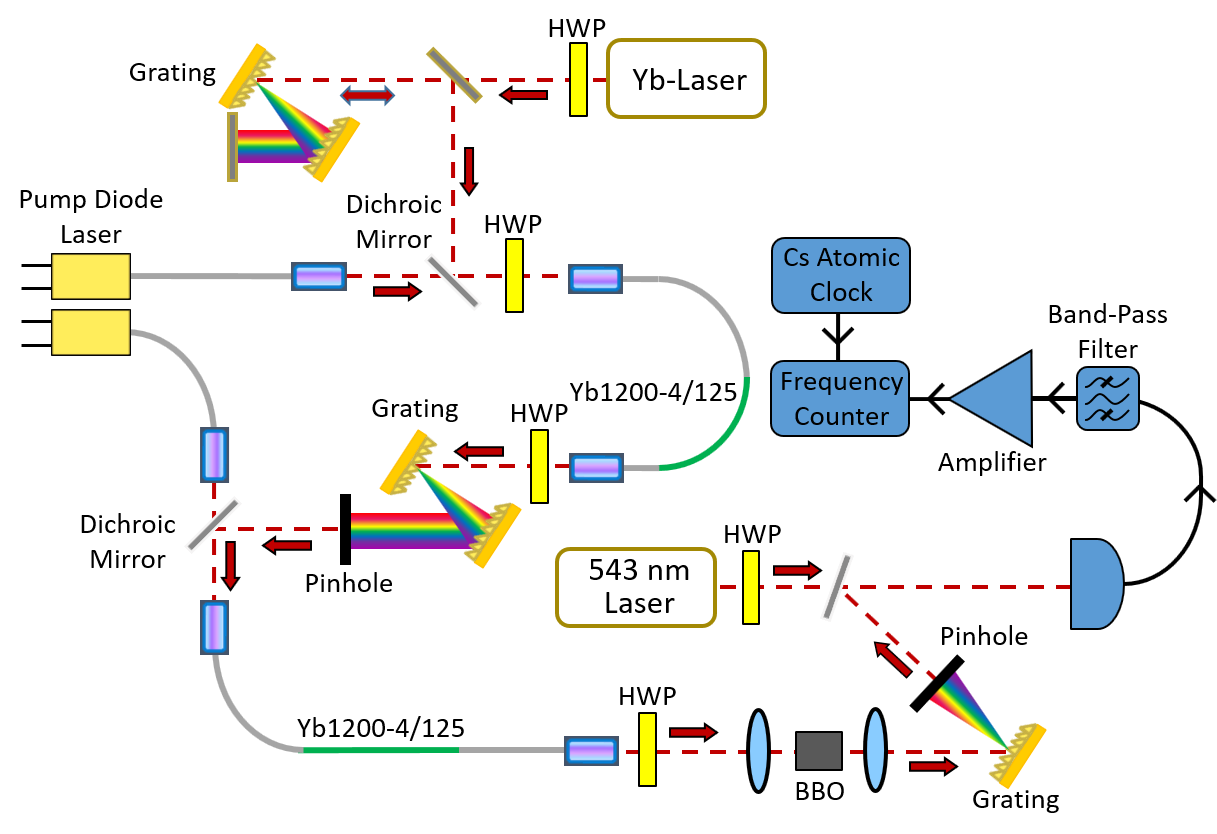}
	\caption{\label{fig:two_mode} The 543-nm He-Ne laser frequency measurement setup. HWP, haf-wave plate.}
\end{figure}

\section{Absolute Frequency Measurements}

As ultimate verification of its stability as well as its accuracy, the frequency comb is used for absolute frequency measurement of a Nd:YAG-${\rm I}_2$ laser that is locked to the ${\rm a}_{10}$ line of the R(56)32-0 group. The frequency of the Nd:YAG laser is locked in the following way: Firstly the second harmonic of the laser frequency is generated with a BBO crystal. The frequency of the Nd:YAG laser is modulated with a piezo at 2 kHz with a modulation amplitude of 1 MHz. Then the standard saturation technique is used and the frequency of the laser is locked with use of the third derivative of the absorption resonance of the ${\rm a}_{10}$ line of R(56)32-0 \cite{NdYAG}. All laser and gas-cell parameters are set to the values recommended by the International Committee for Weights and Measures. After stabilization of the Nd:YAG-${\rm I}_2$ laser, its absolute frequency is measured. For an averaging time of 445 s, the absolute frequency of the 1064-nm output of the Nd:YAG laser is measured as 281 630 111 757 442 $\pm$ 333 Hz, which is within the limits given by the International Committee for Weights and Measures (281 630 111 756 500 $\pm$ 2500 Hz).

As an additional test, we calibrate the absolute frequency of a two-mode stabilized He-Ne laser with wavelength of 543 nm, which is commonly employed for length measurements (Fig. \ref{fig:two_mode}). Pulses from the Yb-fiber laser are compressed with a pair of diffraction gratings and coupled into a fiber amplifier to broaden the laser spectrum to cover 1086 nm. After amplification, the 1086-nm component of the laser spectrum is selected with another pair of diffraction gratings and a pinhole as a bandpass filter. This portion of the spectrum is amplified in another Yb-fiber amplifier. After amplification, the 543-nm comb lines are generated via second-harmonic generation in a BBO crystal. The 543-nm comb lines are further filtered with a diffraction grating and a pinhole. The beat signal between the frequency comb and the He-Ne laser is obtained by the heterodyne beat technique. The beat frequency is counted with a frequency counter. The beat frequency is recorded for 335 s, and the frequency of the laser is measured as 551 580 649 136 940 $\pm$ 7534 Hz. This laser has since been employed in the calibration of numerous length measurements performed at our institute \cite{length_meas}.

\section{Conclusions}

We report a tailored laser oscillator and amplifier design that enables us to obtain very short yet high-energy pulses, while ensuring that the laser operates with net zero cavity dispersion. The pulses are dechirped to 33 fs outside the laser cavity. The noise performance of the laser is characterized through RIN and phase-noise measurements. A stabilized frequency comb that covers the spectral region from $700$ to $1400$ nm is produced with this laser and used for absolute-frequency measurement of a stabilized Nd:YAG-${\rm I}_2$ laser and a 543-nm two-mode stabilized He-Ne laser. Given the practicality of the setup, comprising only readily available optical components, other researchers can easily duplicate and use this simple and robust source.

\section{Acknowledgements}

This work was financially supported by TÜBİTAK 1001 Project No. 109T350, European Research Council Consolidator Grant No. ERC-617521 NLL, and the European Metrology Programme for Innovation and Research (EMPIR) Project No. 15SIB03-OC18. This project has received funding from the EMPIR Programme co-financed by the Participating States and from the European Union’s Horizon 2020 Research and Innovation Programme. 

\bigskip


\bibliography{sample}
\end{document}